\documentclass[smallextended]{svjour3} 

\usepackage[dvips]{graphicx}
\usepackage{amsmath}
\usepackage{amssymb}
\usepackage{color}

\newcommand{\ket}[1]{\left\vert#1\right\rangle}
\newcommand{\bra}[1]{\left\langle#1\right\vert}

\DeclareMathOperator{\Tr}{Tr}

\journalname{Quantum Information Processing}

\begin{document}

\title{Quantum teleportation between a single-rail single-photon qubit and a coherent state qubit
	   using hybrid entanglement under decoherence effects}

\author{Hyunseok Jeong \and    Seunglee Bae \and Seongjeon Choi   }

\institute{Hyunseok Jeong \and  Seunglee Bae \and Seongjeon Choi
			\at Center for Macroscopic Quantum Control, Department of Physics and Astronomy,
				Seoul National University, Seoul 151-742, Korea \\
	%		\email{h.jeong37@gmail.com}
	}

\date{Received: date / Accepted: date}

\maketitle

\begin{abstract}
We study quantum teleportation between two different types of optical qubits using hybrid entanglement as a quantum channel under  decoherence effects.
One type of qubit employs the vacuum and single photon states for the basis, called a single-rail single-photon qubit, and the other utilizes coherent states of opposite phases.
We find that teleportation from a single-rail single-photon qubit to a coherent state qubit is better than the opposite direction in terms of fidelity and success probability.
We compare our results with those using a different type of hybrid entanglement between a polarized single-photon qubit and a coherent state.

\keywords{Quantum teleportation \and Quantum information processing \and Optical qubit}
\PACS{03.67.Hk \and 42.50.Ex}
\end{abstract}

\section{Introduction}
There are a number of possible approaches based on optical systems to quantum information processing.
A well known method is to use single photons as quantum information carriers.
In this type of method, quantum information is encoded in the polarization degree of freedom of a single photon \cite{Single-HV,KLM}, or alternatively, presence and absence of a single photon is used for qubit encoding \cite{Single-VS,LundRalph2002}.
Another possible method utilizes coherent states with opposite phases as a qubit basis \cite{Coherent,Enk2001,Coherent-Tele2,Coherent2,Coherent3,cXZ2}.
Both the approaches have their own advantages and disadvantages for quantum information processing \cite{Single,RalphPryde,SvsC,Hoyong}.
One notable merit of the method based on coherent states is that the Bell-state measurement, a crucial element for optical quantum information processing, can be performed in a nearly deterministic manner using a beam splitter and photodetectors \cite{Coherent-Tele2,JeongPuri}.

Recently, a hybrid approach that combines the advantages of both the methods was proposed, where hybrid entanglement between a single photon with the polarization degree of freedom and a coherent state in a free-travelling field is used as a resource \cite{Hybrid-Genr}.
It enables one to perform a nearly deterministic quantum teleportation and a universal set of gate operations for quantum computing using linear optics and photon detection \cite{Hybrid-CC}.
This type of entanglement is also useful for quantum key distribution and security analysis \cite{Hybrid-Key2}, and has an advantage in performing a loophole-free Bell inequality test using inefficient detectors \cite{Hybrid-Bell}.
In fact, various types of optical hybrid approaches to quantum information processing have been investigated so far in order to find efficient encoding, communication, computation and detection methods \cite{Hybrid-CC,Hybrid-Key2,Hybrid-Bell,Hybrid,Hybrid2,Kimin,Sheng13,Hybrid3,Loock}.

As an application of hybrid entanglement, Park \textit{et al.} studied quantum teleportation between a polarized single-photon qubit and a coherent state qubit \cite{Kimin}.
For this application, hybrid entanglement between a polarized single-photon qubit and a coherent state qubit is required.
Even though a cross-Kerr nonlinearity can be used, in principle, to generate this type of hybrid entanglement \cite{hybrid1,hybrid2,hybrid3}, it is extremely demanding to obtain a clean cross-Kerr interaction of sufficient strength \cite{Kerr1,Kerr2,Kerr3}.
A scheme using a pre-arranged coherent state superposition and linear optics elements was suggested \cite{Hybrid-Genr} while preparation of a coherent state superposition with a high fidelity \cite{cat-gen} is a difficult part in this scheme.
A more feasible method based on the single-photon-addition technique was proposed and experimentally demonstrated \cite{hybrid4}, where hybrid entanglement was generated between a single-photon single-rail qubit (hereafter, a single-rail qubit) \cite{LundRalph2002} and a coherent state qubit.
In this type of hybrid entanglement \cite{hybrid4}, the vacuum and single-photon states, $\ket{0}$ and $\ket{1}$, are used as the basis instead of the horizontal and vertical polarizations of a single photon, $\ket{H}$ and $\ket{V}$.

Thus we study, in this paper, quantum teleportation between a single-rail qubit and a coherent state qubit using hybrid entanglement between those two types of qubits.
We compare our results with the previous work \cite{Kimin} where quantum teleportation was studied using hybrid entanglement between a polarized single photon and a coherent state.

\section{Hybrid entanglement under decoherence}
We consider a hybrid entangled state of a single-rail qubit and a coherent state qubit:
	\begin{eqnarray}
		\ket{\psi}_{sc}
		= \frac{1}{\sqrt{2}}\, \big( \ket{0}_{s}\ket{\alpha}_{c} + \ket{1}_{s}\ket{-\alpha}_{c} \big),
	\label{channel1}
	\end{eqnarray}
where $\ket{\pm\alpha}$ are coherent states of amplitudes $\pm\alpha$. 
Here, the subscripts $s$ and $c$  stand for the single-rail qubit and the coherent state qubit, respectively, and
$\pm\alpha$ are assumed to be real without loss of generality.
We again point out that this type of entanglement was experimentally demonstrated \cite{hybrid4}.

The decoherence effect on state $\rho$ caused by photon loss  is described by the master equation under the Born-Markov approximation  with a zero-temperature environment \cite{Master} as 
	\begin{eqnarray}
		\frac{\partial\rho}{\partial\tau} & = & \hat{J}\rho + \hat{L}\rho \; \\
		\hat{J}\rho	& = & \gamma\, \Sigma_{i}\, \hat{a}_{i}\, \rho\, \hat{a}^{\dagger}_{i}, \nonumber \\
		\hat{L}\rho	& = & -\frac{\gamma}{2}\, \Sigma_{i} \left( \hat{a}^{\dagger}_{i}\, \hat{a}_{i}\, \rho
		+ \rho\, \hat{a}^{\dagger}_{i}\, \hat{a}_{i} \right), \nonumber		
	\label{mastereq}
	\end{eqnarray}
where $\tau$ is the system-environment interaction time, $\gamma$ is the decay constant determined by the coupling strength of the system and environment, and $\hat{a}_{i}(\hat{a}^{\dagger}_{i})$ is the annihilation (creation) operator for mode $i$.
Throughout this paper, we assume that modes $s$ and $c$ undergo the same decoherence time with the same decay constant $\gamma$.
The formal solution of Eq.~(\ref{mastereq}) is \cite{Master-sol}
	\begin{eqnarray}
		\rho\,(\tau) = \exp{\left[(\hat{J}+\hat{L})\, \tau\right]}\, \rho\,(0),
	\end{eqnarray}		
where $\rho\,(0)$ is the initial density operator.
This leads to a time-dependent expression for the initial hybrid channel $\ket{\psi}_{sc}$ in Eq.~(\ref{channel1}) as
	\begin{eqnarray}
		\rho_{sc} \left( \alpha\, ;\, t \right)
		= \frac{1}{2} & \Big( & \ket{0}_{s}\bra{0}
		\otimes \ket{t\alpha}_{c}\bra{t\alpha} \nonumber\\
		& + & \left\{\, t^{2}\ket{1}_{s}\bra{1} + (1-t^{2})\ket{0}_{s}\bra{0}\, \right\}
		\otimes \ket{-t\alpha}_{c}\bra{-t\alpha} \nonumber\\
		& + & t\, e^{-2\alpha^{2}(1-t^{2})} \ket{0}_{s}\bra{1}
		\otimes \ket{t\alpha}_{c}\bra{-t\alpha} \nonumber\\
		& + & t\, e^{-2\alpha^{2}(1-t^{2})} \ket{1}_{s}\bra{0}
		\otimes \ket{-t\alpha}_{c}\bra{t\alpha} \Big),
	\label{channel1decoherence}
	\end{eqnarray}
where $t=\exp{(-\gamma\tau/2)}$ corresponds to the amplitude decay.
We define the normalized interaction time $r=(1-t^{2})^{1/2}$ for later use which gives values of $r=0$ at $\tau=0$ and $r=1$ at $\tau=\infty$.

\section{Teleportation between a single-rail qubit and a coherent state qubit}
Quantum teleportation enables one to transfer an unknown qubit to a distant place using an entangled channel. 
In order to perform quantum teleportation, the sender needs to perform a Bell-state measurement and the receiver should carry out single-qubit transforms based on the outcome of the Bell-state measurement. In order to reflect feasible conditions, we assume that available resources in addition to hybrid entanglement are passive linear optics elements and photon detection.
In this paper, we use notation $s \to c$ for the teleportation from a single-rail qubit to a coherent state qubit, and $c \to s$ for the teleportation in the opposite direction.

\subsection{Teleportation from a single-rail qubit to a coherent state qubit}
We start with the case of $s \to c$.
The teleportation fidelity $F$ is defined as $F=\bra{\psi_{t}}\rho_{out}\ket{\psi_{t}}$ where $\ket{\psi_{t}}$ is the target state of teleportation and $\rho_{out}$ is the output state after completing the teleportation process.
The input state is in the form of a single-rail qubit as
	\begin{eqnarray}
		\ket{\psi_{t}}_{s} = a \ket{0}_{s} + b \ket{1}_{s},
	\label{single}
	\end{eqnarray}
where $a$ and $b$ are unknown coefficients with the normalization condition $|a|^{2} + |b|^{2} = 1$.
It would then be reasonable to choose the target state in the coherent state basis as 
	\begin{eqnarray}
		\ket{\psi_{t}}_{c} = \mathcal{N} \left( a \ket{t\alpha}_{c} + b \ket{-t\alpha}_{c} \right),
	\label{coherent}
	\end{eqnarray}
where $\mathcal{N} = \left( 1+(ab^{*}+a^{*}b)\exp{(-2t^{2}\alpha^{2})} \right)^{-1/2}$ is the normalization factor required due to the nonorthogonality between the two coherent states, $\ket{t\alpha}$ and $\ket{-t\alpha}$.
It is important to note that in order to reflect the decrease of the coherent state amplitude under the photon loss in the entangled channel as Eq.~(\ref{channel1decoherence}), we set the amplitude of the target state accordingly as $t \alpha$.
In this way, we can analyze our system under investigation in a $2\otimes2$-dimensional ``dynamic'' Hilbert space as done in Ref.~\cite{Coherent-Tele2}.

The Bell-state discrimination for the input state (denoted by $s$) with the single-rail qubit part of the channel (denoted by $s'$) are an essential part of quantum teleportation.
The four Bell states are
	\begin{eqnarray}
		\ket{B_{1,2}}_{ss'}
		= \frac{1}{\sqrt{2}}\, \left( \ket{0}_{s}\ket{0}_{s'} \pm \ket{1}_{s}\ket{1}_{s'} \right), \\
		\ket{B_{3,4}}_{ss'}
		= \frac{1}{\sqrt{2}}\, \left( \ket{0}_{s}\ket{1}_{s'} \pm \ket{1}_{s}\ket{0}_{s'} \right).
	\end{eqnarray}
After passing through a 50:50 beam splitter, the Bell states are changed as follows:
	\begin{eqnarray}
		\ket{B_{1}}_{ss'} & \to & \frac{1}{2}\, \left( \ket{2}_{s} \ket{0}_{s'}
		+ \sqrt{2} \ket{0}_{s} \ket{0}_{s'} - \ket{0}_{s} \ket{2}_{s'} \right), \\
		\ket{B_{2}}_{ss'} & \to & \frac{1}{2}\, \left( -\ket{2}_{s} \ket{0}_{s'}
		+ \sqrt{2} \ket{0}_{s} \ket{0}_{s'} + \ket{0}_{s} \ket{2}_{s'} \right), \\
		\ket{B_{3}}_{ss'} & \to & \ket{1}_{s} \ket{0}_{s'}, \\
		\ket{B_{4}}_{ss'} & \to & \ket{0}_{s} \ket{1}_{s'}.
	\end{eqnarray}
As a result, two of the Bell states, $\ket{B_{3}}_{ss'}$ and $\ket{B_{4}}_{ss'}$, can be discriminated using two single-photon detectors at the output modes of the beam splitter.
On the other hand, the other two Bell states cannot be distinguished using linear optics elements \cite{Single-VS,Cal2001,dense}.

The net effect of the Bell-state discrimination of the input state and channel state is equivalent to taking the inner product of the total density operator $\ket{\psi_{t}}_{s}\bra{\psi_{t}} \otimes \rho_{s'c} \left( \alpha\, ;\, t \right)$ with a Bell state.
For example, when one of the Bell states, $\ket{B_{1}}_{ss'}$ , is measured, the output state for the teleportation is
	\begin{eqnarray}
		\rho_{out}^{s \to c}
		= \frac{_{ss'}\bra{B_{1}} \left\{ \ket{\psi_{t}}_{s}\bra{\psi_{t}}
		\otimes \rho_{s'c} (\alpha\, ;\, t ) \right\} \ket{B_{1}}_{ss'}}
		{\Tr{[\ket{B_{1}}_{ss'}\bra{B_{1}} \left\{ \ket{\psi_{t}}_{s}\bra{\psi_{t}}
		\otimes \rho_{s'c} (\alpha\, ;\, t ) \right\}]}},
	\label{out}
	\end{eqnarray}
where the denominator is for normalization.

An appropriate local single-qubit transform  is then applied to the output state in order to reconstruct the target state.
In the case considered above with $\ket{B_{1}}_{ss'}$, no additional operation is required.
The required transforms for the coherent state part for the other cases are
	\begin{eqnarray}
		X_{c} & : & \ket{\pm t \alpha}_{c} \to \ket{\mp t \alpha}_{c}, \nonumber\\
		Z_{c} & : & \ket{\pm t \alpha}_{c} \to \pm \ket{\pm t \alpha}_{c},
	\end{eqnarray}
where $X_{c}$ and $Z_{c}$ correspond to the bit and phase flip operations, respectively, for coherent state qubit. 
When the measurement outcome is $\ket{B_{2}}_{ss'}$, $Z_{c}$ is should be performed, and when $\ket{B_{3}}_{ss'}$ is measured, $X_{c}$ should be applied. For the case of $\ket{B_{4}}_{ss'}$, both $X_{c}$ and $Z_{c}$ are needed.
Since the two of the Bell states that can be identified using linear optics are $\ket{B_{3}}_{ss'}$  and $\ket{B_{4}}_{ss'}$, both $Z_{c}$ and $X_{c}$ are required. We note that $Z_{c}$ is not unitary unless $\alpha\rightarrow\infty$.  

It is straightforward to perform $X_{c}$ using a phase shifter described by $\exp{(i \varphi\, \hat{a}^{\dagger} \hat{a})}$ with $\varphi=\pi$.
On the other hand, the implementation of $Z_{c}$ is not straightforward due to the nonorthogonality between the coherent states, but there are several possible methods.
The simplest way is to use the displacement operation, which is a unitary transform, in order to approximate $Z_{c}$ for relatively large values of $\alpha$ \cite{Coherent2}.
This can be performed using a strong coherent field and a beam splitter with a high transmissivity.
Another possible method is to use an additional teleportation circuit for the coherent state qubit via an entangled coherent state as the quantum channel \cite{Coherent3}.
Since quantum teleportation without the single-qubit  transforms makes the input qubit bit-flipped or phase-flipped depending on the Bell-state measurement, one can perform the phase-flip operation with success probability of 1/2 together with the $X_c$ operation.
The teleportation process can be applied successively until the phase flip operation is obtained.
However, this method requires entangled coherent states and detectors as additional resource \cite{Coherent3}.
It is worth noting that when quantum teleportation is used for quantum computing, the single-qubit operations  
may not be necessary because they can be absorbed into the error correction process via the Pauli frame technique \cite{Knill2005}.

Inserting the explicit form of $\rho_{s'c} \left( \alpha\, ;\, t \right)$ in Eq.~(\ref{channel1decoherence}) and $\ket{\psi_{t}}_{s}$ in Eq.~(\ref{single}) into Eq.~(\ref{out}), we get
	\begin{eqnarray}
		\rho_{out}^{s \to c}
			& = & \mathcal{M}\, \Big\{ \vert a \vert^{2} \ket{t\alpha}\bra{t\alpha}
			+ \left[ (1-t^{2}) \vert a \vert^{2} + t^{2} \vert b \vert^{2}\, \right]
			\ket{-t\alpha}\bra{-t\alpha} \nonumber \\
			& & +\, t\, e^{-2 \alpha ^{2} (1-t^{2})}
			\left[\, a b^{*} \ket{t\alpha}\bra{-t\alpha}+ a^{*} b \ket{-t\alpha}\bra{t\alpha}\, \right]
			\Big\},
	\label{scout}
	\end{eqnarray}
where $\mathcal{M} = \left\{ (2-t^{2})\vert a \vert^{2} + t^{2}\, \vert b \vert^{2} + t\, e^{-2 \alpha ^{2}} (a b^{*} + a^{*} b) \right\}^{-1}$.
The fidelity between the output state $\rho_{out}^{s \to c}$ in Eq.~(\ref{scout}) and the target state $\ket{\psi_{t}}_{c}$ in Eq.~(\ref{coherent}) is
	\begin{eqnarray}
		F_{s \to c}
		& = &\, _{c}\bra{\psi_{t}} \rho_{out}^{s \to c} \ket{\psi_{t}}_{c} \nonumber \\
		& = & \mathcal{N}^{2} \mathcal{M} \times \Big\{
		\big\vert a\, ( a + b\, e^{-2 t^{2} \alpha ^{2}} ) \big\vert^{2}
		+ \left[ (1-t^{2}) \vert a \vert^{2} + t^{2} \vert b \vert^{2} \right]
		\big\vert ( a\, e^{-2 t^{2} \alpha ^{2}}   + b ) \big\vert^{2} \nonumber \\
		& & + 2\, t\, e^{-2 \alpha^{2} (1-t^{2})}\,
		\mathrm{Re}\, \big[ a\, b^{*} \left( a\, e^{-2 t^{2} \alpha ^{2}} + b \right)
		\left( a^{*} + b^{*}\, e^{-2 t^{2} \alpha ^{2}} \right) \big] \Big\}.
	\end{eqnarray}
We need the average fidelity over all possible input states.
It can be found by parametrizing the coefficients of the input state as $a=\cos\, [\, \theta /2\, ] \exp[\, i\, \phi /2\,]$ and $b=\sin\, [\, \theta /2\, ] \exp[\, -i\, \phi /2\,]$, where $0 \leq \theta < \pi$ and $0 \leq \phi < 2\pi$.
The formal expression of the average fidelity $F_{s \to c}^{ave}$ is
	\begin{eqnarray} 
		F_{s \to c}^{ave} = \frac{1}{4\pi} \int_{0}^{\pi} \mathrm{d} \theta \ \sin{\theta}
		\int_{0}^{2\pi} \mathrm{d} \phi \ F_{s \to c}.
	\label{avefid}	
	\end{eqnarray}
We have numerically performed the integration in Eq.~(\ref{avefid}) for several values of $\alpha$, and plot the results in Fig.~\ref{Fidelity}.

\subsection{Teleportation from a coherent-state  qubit to a single-photon qubit}
We now consider the case of $c \to s$, where the input qubit is a coherent state qubit in the form of Eq.~(\ref{coherent}) and the target state is a single-rail qubit in Eq.~(\ref{single}).
The Bell-state measurement for coherent state qubits can be performed by using a 50:50 beam splitter and two photon-number parity measurements \cite{Coherent-Tele2}.
The four Bell states in the dynamic coherent state basis are
	\begin{eqnarray}
		\ket{\mathcal{B}_{1,2}}_{cc'}
		& = & \mathcal{N}_{\pm} \left( \ket{t\alpha}_{c}\ket{t\alpha}_{c'} \pm
		\ket{-t\alpha}_{c}\ket{-t\alpha}_{c'} \right), \nonumber \\
		\ket{\mathcal{B}_{3,4}}_{cc'}
		& = & \mathcal{N}_{\pm} \left( \ket{t\alpha}_{c}\ket{-t\alpha}_{c'} \pm
		\ket{-t\alpha}_{c}\ket{t\alpha}_{c'} \right),
	\end{eqnarray}
where $\mathcal{N}_{\pm} = \left( 2 \pm 2 \exp(-4t^2\alpha^2) \right)^{-1/2}$ are normalization factors.
The Bell states evolve through the 50:50 beam splitter as 
	\begin{eqnarray}
		\ket{\mathcal{B}_{1}}_{cc'} & \to \mathcal{N}_{+} \ket{even}_{c} \ket{0}_{c'}, \qquad
		\ket{\mathcal{B}_{2}}_{cc'} & \to \mathcal{N}_{-} \ket{odd}_{c} \ket{0}_{c'}, \nonumber \\
		\ket{\mathcal{B}_{3}}_{cc'} & \to \mathcal{N}_{+} \ket{0}_{c} \ket{even}_{c'}, \qquad
		\ket{\mathcal{B}_{4}}_{cc'} & \to \mathcal{N}_{-} \ket{0}_{c} \ket{odd}_{c'},
	\end{eqnarray}
where $\ket{even}= \ket{\sqrt{2}\, t \alpha} + \ket{-\sqrt{2}\, t \alpha}$ has nonzero photon-number probabilities only for even numbers of photons and $\ket{odd}= \ket{\sqrt{2}\, t \alpha} - \ket{-\sqrt{2}\, t \alpha}$ has nonzero photon-number probabilities only for odd numbers of photons.
The parity measurement projection operators $O_{j}$,
	\begin{eqnarray}
		\hat{O_{1}}&=&\sum_{n=1}^{\infty} \ket{2n}_{c}\bra{2n} \otimes \ket{0}_{c'}\bra{0}, \nonumber \\
		\hat{O_{2}}&=&\sum_{n=1}^{\infty} \ket{2n-1}_{c}\bra{2n-1} \otimes \ket{0}_{c'}\bra{0}, \nonumber \\
		\hat{O_{3}}&=&\sum_{n=1}^{\infty} \ket{0}_{c}\bra{0} \otimes \ket{2n}_{c'}\bra{2n}, \nonumber \\
		\hat{O_{4}}&=&\sum_{n=1}^{\infty} \ket{0}_{c}\bra{0} \otimes \ket{2n-1}_{c'}\bra{2n-1},
	\end{eqnarray}
where subscript $j$ corresponds to the $j$-th Bell state, can be used to discriminate between the four states.
It should be noted that there is a nonzero probability of getting $\ket{0}_{c}\ket{0}_{c'}$ for which neither of the detectors registers any photon.
Such as a case is regarded as a failure event, and the failure probability is $P_f=\exp\,[-2 t^2\alpha^2]$ \cite{Coherent-Tele2,Hybrid-CC}.
We shall further discuss the success probability of the teleportation process later in this paper.

According to the standard teleportation protocol, when $\ket{\mathcal{B}_{1}}_{cc'}$ is measured, no additional operation is required.
The output state with the normalization factor is
	\begin{eqnarray}
		\rho_{out}^{c \to s}
		& = &\, \frac{\Tr_{cc'} \left\{ (\hat{O}_{1})_{cc'} (\hat{U}_{BS})_{cc'}
		\big[\, \rho_{sc'} \left( \alpha\, ;\, t \right) \otimes \ket{\psi}_{c}\bra{\psi}\, \big]
		(\hat{U}_{BS}^{\dagger}) \right\}}
		{\Tr \left\{ (\hat{O}_{1})_{cc'} (\hat{U}_{BS})_{cc'}
		\big[\, \rho_{sc'}\left( \alpha\, ;\, t \right) \otimes \ket{\psi}_{c}\bra{}\psi\, \big]
		(\hat{U}_{BS}^{\dagger})   \right\}} \nonumber\\
		& = & \, \left( \vert a \vert^{2} + (1-t^{2}) \vert b \vert^{2} \right) \ket{0}\bra{0}
		+ t^{2}\, \vert b \vert^{2} \ket{1}\bra{1} \nonumber\\
		& & +\, t\, e^{-2\alpha^{2} (1-t^{2})}
		\left( a b^{*} \ket{0}\bra{1} + a^{*} b \ket{1}\bra{0} \right),
	\end{eqnarray}
where $\hat{U}_{BS}$ represents the 50:50 beam splitter operator defined as $U_{BS} = \exp \left[\, \frac{\pi}{4}\, (\, \hat{a}^{\dagger}_{0} \hat{a}_{1} - \hat{a}_{1} \hat{a}^{\dagger}_{0}\, )\, \right]$.
If $\ket{{\cal B}_{2}}_{cc'} $ is  measured, the Pauli-Z operation for the single-rail qubit is required to complete the teleportation process, which can be performed by a $\pi$-phase shifter.
When $\ket{{\cal B}_{3}}_{cc'} $ and $\ket{{\cal B}_{4}}_{cc'} $
 are measured, the Pauli-X  operation is needed to implement the bit flip, $\ket{0} \leftrightarrow \ket{1}$, which is difficult to realize using linear optics.
We shall thus take only $\ket{{\cal B}_{1}}_{cc'}$ and $\ket{{\cal B}_{2}}_{cc'}$ as successful Bell-measurement outcomes.
The fidelity between the output state and the target state is 
	\begin{eqnarray}
		F_{c \to s}
		& = &\,  _{s}\bra{\psi_{t}} \rho_{out}^{c \to s} \ket{\psi_{t}}_{s} \nonumber\\
		& = & \vert a \vert^{4} +
		\left( (1-t^{2}) + 2\, t\, e^{-2\alpha^{2} (1-t^{2})} \right)\, \vert a \vert^{2} \vert b \vert^{2}
		+ t^{2}\, \vert b \vert^{4},
	\end{eqnarray}
and its average can be calculated using Eq.~(\ref{avefid}) as
	\begin{eqnarray}
		F_{c \to s}^{ave}
		= \frac{2}{3} + \frac{t^{2}+2\, t\, e^{-2\alpha^{2}(1-t^{2})}}{6}.
	\end{eqnarray}

\begin{figure}[t]
	\centering
	\includegraphics[width=0.88\textwidth]{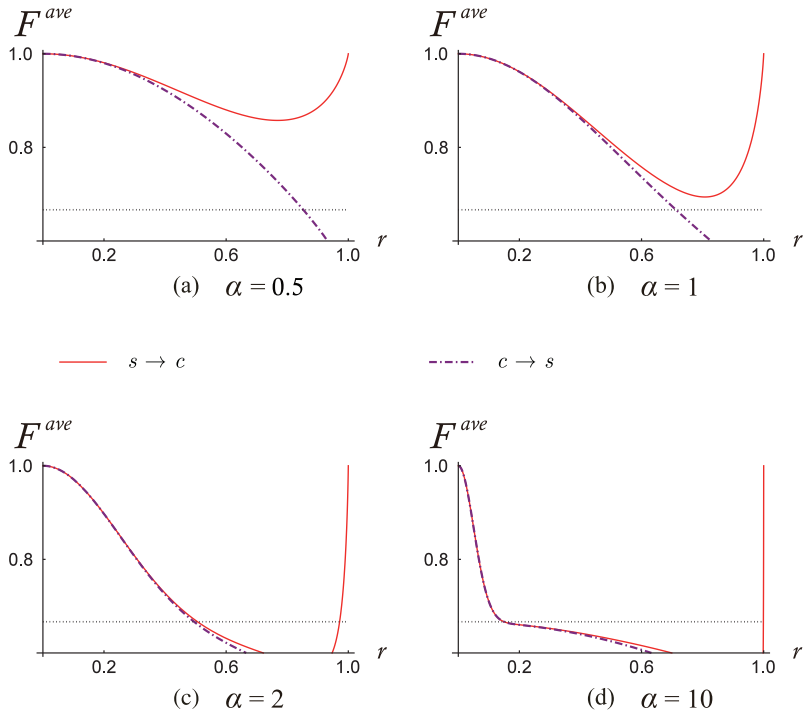}
	\caption{Average fidelities of teleportation as a function of the normalized time $r$ for
	         (a) $\alpha = 0.5$, (b) $\alpha = 1$, (c) $\alpha = 2$ and (d) $\alpha = 10$.
	         The red solid curves correspond to the cases for $s \to  c$ and
	         the purple dot-dashed curves are for $c \to  s$.
	         The horizontal dotted line is the classical limit, 2/3, given for comparisons.}
	\label{Fidelity}
\end{figure}

In Fig.~\ref{Fidelity}, the average teleportation fidelities for both the directions of $s \to c$ and $c \to  s$ are plotted against the normalized time $r$ for several different amplitudes.
A conspicuous observation is that the fidelity for the direction of $s \to c$ is always higher than that of the other direction regardless of the values of $\alpha$ and $r$.
However, as we shall see in Sec.~\ref{Sec4}, this gap between the fidelities for the two directions is smaller than the gap when the other type of hybrid entanglement \cite{Kimin} between a coherent state qubit and a polarized single-photon qubit is employed.
The teleportation fidelities for both  $s \to c$ and $c \to s$ decrease more rapidly as $\alpha$ becomes larger.
This  is due to the fact that the hybrid entanglement has the properties of a macroscopic superposition when
$\alpha$ is large \cite{hybrid4,review-jeong}.
When $r$ approaches 1, the teleportation fidelity for  $s \to c$ goes up to 1. 
The reason for this is that the target state, $\ket{\psi_t}_c = \mathcal{N}\, \left( a\, \ket{t\alpha} + b\, \ket{-t\alpha} \right)$, approaches the vacuum in this limit and the overlap between the target state and the classical mixture of $\ket{t\alpha}$ and $\ket{-t\alpha}$, $\mathcal{N}^2 \left(\, |a|^2\, \ket{t\alpha }\bra{t\alpha} + |b|^2\, \ket{-t\alpha}\bra{-t\alpha}\, \right)$, approaches 1 accordingly.

\subsection{Success Probabilities}
When using linear optics and photodetectors, the success probability of the Bell-state measurement is limited and certain required single-qubit transforms are unavailable.
These factors make the success probability of the teleportation process also to be limited.

In the case of $s \to c$, the Bell-state measurement for single-rail qubits can identify only two of the four Bell-states.
We pointed out that one of the local transforms, $Z_c$, are non-trivial but there are a couple of possible methods to implement it.
Considering the inherent limitation of the Bell measurement, the success probability of teleportation for $s \to c$ can be considered to be $1/2$ when there is no photon loss, i.e., when $r=0$.
The photon loss process causes some of the qubit elements in state $\ket{1}$ to evolve to $\ket{0}$.
This type of loss cannot be noticed by photodetectors used for the Bell-state measurement because any decohered single-rail qubit remains within the original two-dimensional qubit space. 
 As an extreme example, 
let us suppose that the channel is fully decohered for $r\rightarrow\infty$, i.e., it has become the vacuum.
In this case, the two modes for the Bell-state measurement can be represented as
	\begin{eqnarray}
		& & \left( a\ket{0}_{s} + b\ket{1}_{s} \right) \ket{0}_{s'} \nonumber\\
		& & = \frac{a}{\sqrt{2}} \left( \ket{B_{1}}_{ss'} + \ket{B_{2}}_{ss'} \right) +
		\frac{b}{\sqrt{2}} \left( \ket{B_{3}}_{ss'} - \ket{B_{4}}_{ss'} \right).
	\end{eqnarray}
Note that $ \ket{B_{3}}$ and  $\ket{B_{4}}$ correspond to successful events while the other two Bell states cannot be identified.
It is straightforward to notice that taking average over $a$ and $b$ for the input state, the success probability 
of the teleportation process is $P_{s \to c}^{ave}=1/2$ for $r\rightarrow\infty$.
In fact, no matter whether the single photon $|1\rangle_{s'}$ or the vacuum $|0\rangle_{s'}$ is incoming for mode $s'$, the success probability does not change; this means that  $P_{s \to c}^{ave}=1/2$  regardless of the value of the decoherence time $r$. 

\begin{figure}[t]
	\centering
	\includegraphics[width=0.55\textwidth]{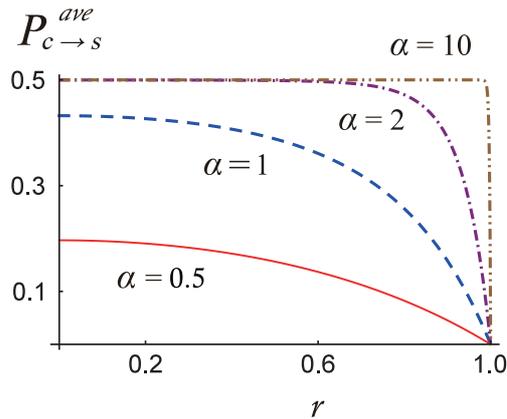}
	\caption{Success probabilities for teleportation from a coherent-state qubit to a single-rail qubit ($c \to s$)
			 with several values of amplitudes ($\alpha = 0.5, 1, 2, 10$) against the normalized time $r$.
			 We also note that the success probability for the case of $s \to c$ is 1/2 regardless of the value of $r$, which is not plotted.
			 }
	\label{SucProb}
\end{figure}

The Bell-state measurement for coherent state qubits, required for the case of $c \to s$, can identify all four Bell states with the success probability of $1 - e^{-2  t^{2} \alpha^{2}}$ \cite{Coherent-Tele2,Hybrid-CC}.
However, we pointed out that a  local single-qubit operation, the Pauli-X operation which flips $\ket{0}$ and $\ket{1}$, cannot be effectively performed using linear optics elements.
We thus take only two outcomes of the Bell-state measurements as successful events and the average success probability the teleportation in this case is
	\begin{eqnarray}
		P_{c \to s}^{ave} = \frac{1 - e^{-2 t^{2} \alpha^{2} }}{2}.
	\end{eqnarray}
Clearly, $P_{s \to c}^{ave}$ is always larger than $P_{c \to s}^{ave}$ but they become identical in the limit of $t\alpha\gg1$.
We plot the teleportation success probabilities for several values of $\alpha$ in Fig.~\ref{SucProb} which 
shows that $P_{c \to s}^{ave}$ becomes 1/2 as $\alpha$ increases.

\section{Comparison between two different hybrid schemes}
\label{Sec4}
In this section, we compare our results with the previous study in Ref.~\cite{Kimin}, where the quantum teleportation between a photon-polarization qubit and a coherent state qubit (hereafter denoted as $p \leftrightarrow c$) was studied.
The quantum channel used for teleportation between a photon-polarization qubit and a coherent state qubit \cite{Kimin} is a hybrid entangled state in the form of 
	\begin{eqnarray}	
		\ket{\psi}_{sc}
		= \frac{1}{\sqrt{2}}\, \left( \ket{H}_{p}\ket{\alpha}_{c} + \ket{V}_{p}\ket{-\alpha}_{c} \right),
	\label{channel2}
	\end{eqnarray}
where subscript $p$ represents the polarization qubit.
The input or target state for the polarization qubit is
	\begin{eqnarray}
		\ket{\psi_{t}}_{p}
		= a \ket{H}_{p} + b \ket{V}_{p}.
	\label{polarization}
	\end{eqnarray}

Similarly to the case of single-rail qubits, the Bell-state measurement for polarized single-photon qubits can discriminate  only two of four Bell states using linear optics elements \cite{Cal2001,Lukenhause99} while their single-qubit transforms are straightforward \cite{KLM}.
The results are discarded only when no photons are detected in the Bell-state measurement.
Of course, when loss caused by decoherence occurs, the parity measurement scheme used for the Bell-state measurements in the coherent state basis cannot filter out `wrong results' in the polarization part, which is obviously different from the Bell-state measurement with polarization qubits, and this type of error will be reflected in the degradation of the fidelity.

\begin{figure}[ht]
	\centering
	\includegraphics[width=0.88\textwidth]{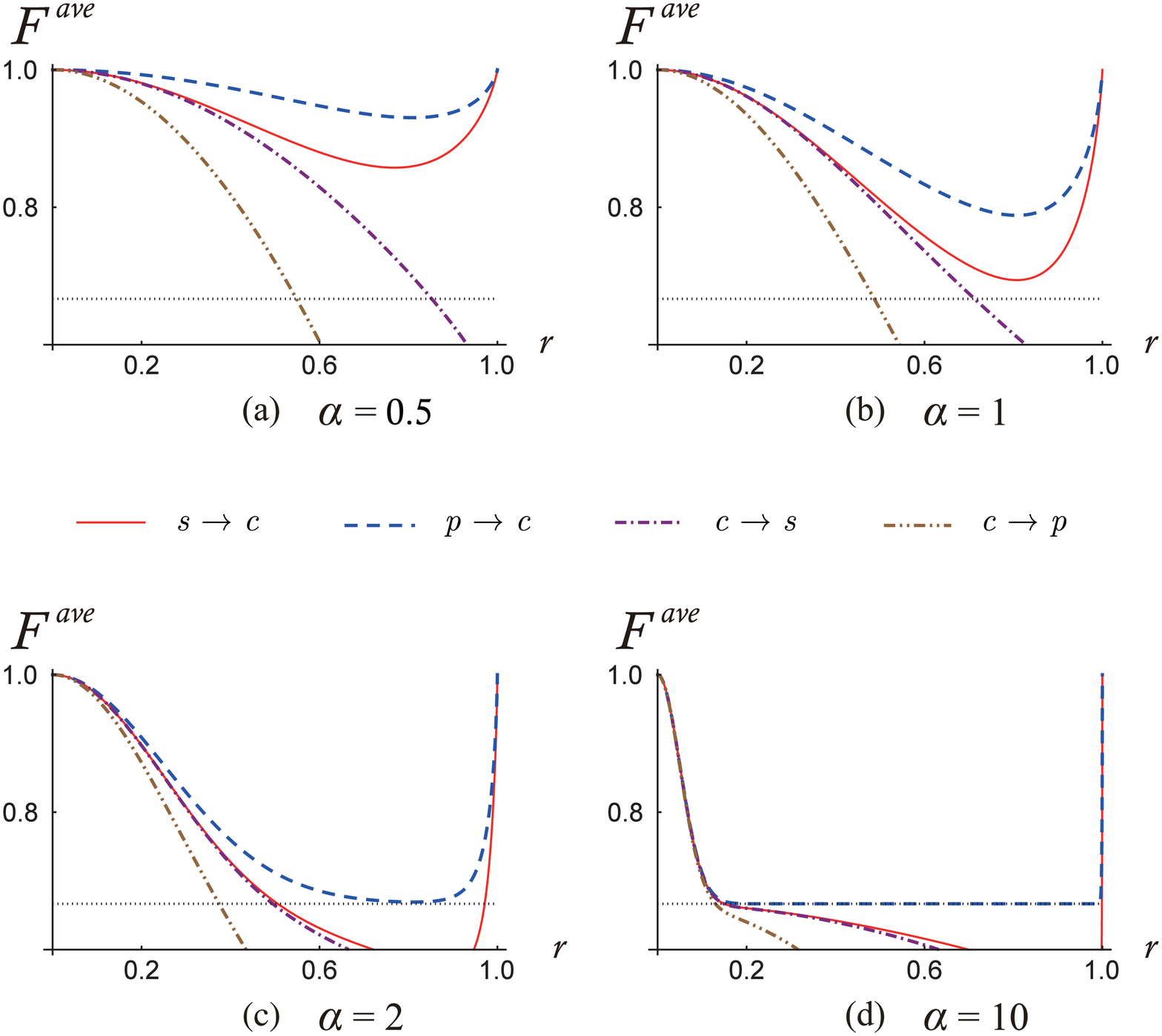}
	\caption{Average fidelities of teleportation of $p \to c$ (blue dashed curve),
			 $s \to c$ (red solid), $c \to s$ (purple dot-dashed)
			 and $c \to p$ (brown doubledot-dashed) for several different values of $\alpha$.
			 The classical limit $2/3$ (horizontal dotted line) is given for comparison.}
	\label{compare}
\end{figure}

The fidelities of the two cases obtained in Ref.~\cite{Kimin} are
	\begin{eqnarray}
		F_{c \to p}
		=  t^{2} \left[\, \vert a \vert^{4} + \vert b \vert^{4}
		+ e^{-2 \alpha^{2} (1-t^{2})}\,\vert a \vert^{2}\, \vert b \vert^{2}\, \right].
	\end{eqnarray}
and
	\begin{eqnarray}
		F_{p \to c}
		 = & & \mathcal{N}^{2}\, \mathcal{S}
		\times\, \Big\{ \big\vert a\, ( a + b\, e^{-2 t^{2} \alpha ^{2}} ) \big\vert^{2}
		+ \big\vert b\, ( a\, e^{-2 t^{2} \alpha^{2}} + b ) \big\vert^{2} \nonumber \\
	    & & +\, 2\, e^{-2 \alpha^{2} (1-t^{2})}\,\,
	    \mathrm{Re}\, \left[\, a\, b^{*} \big( a + b\, e^{-2 t^{2} \alpha ^{2}}  \big)
	    \big( a^{*}\, e^{-2 t^{2} \alpha^{2}} + b^{*} \big) \right] \Big\},
	\end{eqnarray}
where $\mathcal{S} = \left(  1 + e^{-2 \alpha ^{2}} (a b^{*} + a^{*} b) \right)^{-1}$.
Using Eq.~(\ref{avefid}), the average fidelity $F^{ave}_{c \to p} $ is obtained as
	\begin{eqnarray}
		F^{ave}_{c \to p} = t^{2}\, \left( \frac{2 + e^{-2 \alpha^{2} (1-t^{2})}}{3} \right).
	\end{eqnarray}
and $F^{ave}_{p \to c} $ can be numerically calculated for given values of $\alpha$.
The average success probabilities for the two cases are~\cite{Kimin}
	\begin{eqnarray}
		P_{p \to c}^{ave}
		& = & \frac{t^{2}}{2}, \\
		P_{c \to p}^{ave}
		& = & \frac{e^{2 \alpha^{2} t^{2}}-1}{2}\,
		\ln{\frac{1+e^{-2 \alpha^{2} t^{2}}}{1-e^{-2 \alpha^{2} t^{2}}}}.
	\end{eqnarray}

	\begin{figure}[t]
	\centering
	\includegraphics[width=0.88\textwidth]{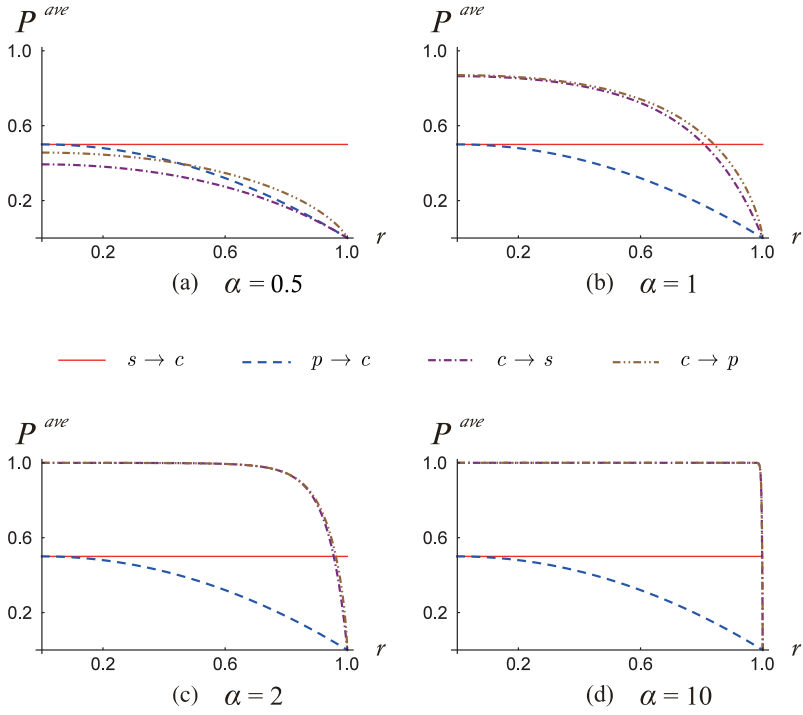}
	\caption{Average success probabilities of $p \to  c$ (blue dashed curve),
			 $s \to c$ (red solid), $c \to s$ (purple dot-dashed)
			 and $c \to p$ (brown double-dot-dashed) for several different values of $\alpha$.}
	\label{compare2}
\end{figure}

Figures~\ref{compare} and \ref{compare2} present the average fidelities and success probabilities for the four cases of teleportation.
In terms of fidelities, the case of $p \to c$ is better than the case if $s \to c$.
The reason is explained as follows. The Bell-state measurement for polarized single-photon qubits allows one to discard failure events whenever photon loss occurs. However, in the case of single-rail qubits, where the vacuum and single photon states form the qubit basis, the photon losses cannot be identified by detectors, which degrades the fidelity of the output state.
Because of the same reason, the success probability for $p \to c$ is lower than that for $s \to c$ (Fig.~4).

On the other hand, the fidelity of  $c \to s$ is better than the $c \to p$ case as shown in Fig.~\ref{compare}. In these cases, the Bell-state measurement for coherent-state qubits are used, where the photon losses cannot be identified by detectors. When photon losses occur in the channel during the teleportation process, the output state for  $c \to s$ remains in the qubit space of the vacuum and single photon. However, when photons are lost in the channel, the output state for $c \to p$ would contains the vacuum state in addition to $\ket{H}$ and $\ket{V}$, which degrades the fidelity more. The success probabilities for the two cases are close to each other as shown in Fig.~\ref{compare2}.

\section{Remarks}
We have investigated quantum teleportation between two distinct types of optical qubits under photon loss effects.
One type of qubit is of the vacuum and single photon as the basis states, called a single-rail qubit, and the other is of coherent states of opposite phases.
First, the average fidelity of teleportation from a single-rail qubit to a coherent state qubit ($s\rightarrow c$) is always larger than the opposite direction ($c\rightarrow s$) under photon loss. This is due to the fact that failure events caused  by  photon loss are always noticed by the detectors for the Bell-sate measurement and they can be discarded whenever they occur. This enhances the fidelity for the former case.
It should be noted that the non-orthogonality of the two coherent states are another factor which increases the fidelity for $c\rightarrow s$ when $\alpha$ is small.

The success probability of teleportation from a coherent-state qubit to a single-rail qubit becomes higher up to 1/2 as the coherent amplitude $\alpha$ gets larger, while it is reduced by the  decoherence time.
This is due to Bell-state discrimination for coherent state can be done perfectly for high $\alpha$.
In the case of the opposite direction ($s \rightarrow c$), however, the success probability of teleportation is 1/2 regardless of the decoherence time and the coherent amplitude because only two of the four Bell states in terms of single-rail qubits can be discriminated with equal rages at any condition.

We have further compared our result with a previous related study using another type of hybrid entanglement between a coherent-state qubit and a polarized single-photon qubit \cite{Kimin}.
The average fidelity of teleportation from a polarization qubit to a coherent state qubit ($p \rightarrow c$) is found to be always larger than that from a single-rail qubit to a coherent state qubit $s \rightarrow c$.
Meanwhile, the average fidelity of teleportation for $c \rightarrow s$ is larger than that for $c \rightarrow p$.
In terms of the success probability, the teleportation from
a single-rail qubit to a coherent state qubit is always better than that from  a polarized single-photon qubit to a coherent-state qubit.
These can be attributed to the difference between the Bell-state measurement for polarized single-photon qubits  and that for single-rail qubits; the former can filter out failure events due to photon losses while the latter cannot do so.
On the other hand, the fidelity of  $c \to s$ is better than the case of $c \to p$. 
This is due to the difference between the decoherence mechanism of single-rail qubits and that of polarized single-photon qubits; while the former remains in the Hilbert space of the vacuum and single photon states even after any amount of decoherence, the latter gets out of  the original Hilbert space by photon loss effects.

Our study reveals, in detail, advantages and disadvantages of different types teleportation schemes using hybrid entanglement of light for efficient quantum information processing based on hybrid architectures of optical systems.

\section{Acknowledgements}
This work was supported by the National Research Foundation of Korea (NRF) through a grant funded by the Korean government (MSIP) (Grant No. 2010-0018295).
%The authors thank useful discussions with H. Kwon.

\end{document}